# Earth wind as a possible source of lunar surface hydration

H. Z. Wang[1,2], J. Zhang[1*], Q. Q. Shi[1,2*], Y. Saito[3], A. W. Degeling[1], I. J. Rae[4], J. Liu[2], R. L. Guo[5], Z. H. Yao[6], A. M. Tian[1], X. H. Fu[1], Q.G. Zong[7], J. Z. Liu[8], Z. C. Ling[1], W. J. Sun[9], S. C. Bai[1], J. Chen[1], S. T. Yao[1], H. Zhang[5], Y. Wei[5], W. L. Liu[10], L. D. Xia[1], Y. Chen[1], Y. Y. Feng[2], S. Y. Fu[7] and Z. Y. Pu[7]

[1]Shandong Provincial Key Laboratory of Optical Astronomy and Solar-Terrestrial Environment, Institute of Space Science, Shandong University, Weihai, Shandong, China.

[2]State Key Laboratory of Space Weather, National Space Science Center (NSSC), Chinese Academy of Sciences, Beijing, China.

[3]Institute of Space and Astronautical Science, Japan Aerospace Exploration Agency, Sagamihara, Japan.

[4]Mullard Space Science Laboratory, University College London, Dorking, UK.

[5]Key Laboratory of Earth and Planetary Physics, Institute of Geology and Geophysics, Chinese Academy of Sciences, Beijing, China.

[6]Laboratoire de Physique Atmosphérique et Planétaire, STAR Institute, Université de Liège, Liège, Belgium.

[7]School of Earth and Space Sciences, Peking University, Beijing, China.

[8]Center for Lunar and Planetary Sciences, Institute of Geochemistry, Chinese Academy of Sciences, Guiyang, China.

[9]Department of Climate and Space Sciences and Engineering, University of Michigan, Ann Arbor, US.

[10]School of Space and Environment, Beihang University, Beijing, China.

**Abstract:** Understanding the sources of lunar water is crucial for studying the history of lunar evolution, and also the solar wind interaction with the Moon and other airless bodies. Recent observations revealed lunar hydration is very likely a surficial dynamic process driven by solar wind. Solar wind is shielded over a period of 3-5 days as the Moon passes through the Earth's magnetosphere, during which a significant loss of hydration is expected from previous works. Here we study lunar hydration inside the magnetosphere using orbital spectral data, which unexpectedly found that the polar surficial OH/$H_2O$ abundance remains at the same level when in the solar wind and in the magnetosphere. We suggest that particles from the magnetosphere (Earth wind, naturally different from solar wind) contribute to lunar hydration. From lunar orbital plasma observations, we find the existence of optimal energy ranges, other than 1 keV as previously thought, for surface hydration formation. These optimal energy ranges deduced from space observations may provide strong implications for laboratory experiments simulating lunar hydration processes.

**Main Text:** Recent orbital observations and laboratory analyses revealed a "wet" Moon with various forms of lunar water (*1-10*). These forms of water have been proposed to come from both indigenous and exogenous sources, including lunar interior, comets, asteroids, solar wind, and the Earth's magnetosphere (*11-13*). Therefore, understanding the sources and formation processes of lunar water is key for studying the origin and evolution of planetary bodies, including lunar magmatic evolution, bombardment history, and solar wind-surface interaction on the Moon and other airless bodies.

Evidence of water from the lunar interior was found by laboratory analyses of lunar samples, such as pyroclastic glasses, lunar melt inclusions, apatite grains, and anorthosites (*1*, *9*, *10*, *14*, *15*), and was also detected in pyroclastic deposits by orbital observations (*16*). In contrast, examples of exogenous origins include the episodic delivery of water to the Moon from meteoroids and comets (*17*), which result in deuterium/hydrogen (D/H) values (*8*, *18*) that differ from lunar interior sources.

Solar wind hydrogen is considered to be another exogenous source of water (*2-4*, *13*, *19-27*). Surface water content has been found to exhibit diurnal variations, which is interpreted to indicate a dynamic balance between a continuous source from the solar wind and loss processes dependent on solar illumination and surface temperature (*2-4*, *20*, *24-26*). Lunar soil analysis and laboratory ion irradiation experiments provide further evidence that solar wind hydrogen provides an exogenous source (*28-31*). The interpretation from these studies is that lunar water is generated by solar wind hydrogen implantation into the upper-most surface of lunar soil grains and forming OH bonds and even $H_2O$ (*13*, *19*).

During about three quarters of the lunar orbit, the Moon is immersed in solar wind that provides a predominant proton flux at ~1 keV, with 4 % comprised of alpha particles and other heavy ions (*32*). During the remaining 3-5 days of every lunation, the Moon lies within the Earth's magnetosphere (Fig. 1A), in which most of the solar wind particles are shielded. In the absence of other hydrogen sources, a decrease of hydration would be expected due to surface loss processes such as thermal diffusion and photodissociation (*12*, *13*). However, the Earth's magnetosphere is not empty, instead occupied by the Earth wind whose ion constituents and energies are different from the solar wind. The particle species in the Earth wind consist of both solar wind and terrestrial species (i.e., $H^+$, $He^+$, $O^+$, $N_2^+$, $NO^+$, $O_2^+$) (*33-39*). Particle fluxes are typically several times lower than that in the solar wind (*40*), however the particle energy distribution is significantly broader. Ions from Earth wind have been detected in the vicinity of the Moon (*41*).

Therefore, the first questions that arise are whether particles from Earth wind can reach the surface of the Moon, and whether they can contribute to lunar surface hydration as an exogenous source, similar to the solar wind. Attempting to answer these questions, we studied the spatial distribution and temporal variations of lunar water at high latitudes with the Chandrayaan-1 Moon Mineralogy Mapper ($M^3$) data, comparing and contrasting intervals when the Moon lies inside/outside the magnetosphere.

**Results:** The maps of lunar surficial hydration at polar regions are presented in Fig.1 (B-C). In general, we find that there are no significant differences in OH/$H_2O$ abundance for intervals when the Moon is located in the Earth's magnetosphere and in the solar wind. Then, we quantitatively studied the spatial and temporal variations of lunar surface hydration, especially those inside/outside the Earth's magnetosphere using epoch analysis. The full-moon time, around which the Moon was within the magnetosphere, are taken as zero epoch, with the previous and next new moon times corresponding to the left- and right-hand sides. The data points are binned into 5° latitude and 24 Earth hour intervals. As shown by Fig. 1 (B-C) and 2 (A-B), the lunar surface OH/$H_2O$ abundance increases with latitude toward the polar regions, which is consistent with previous studies (*4*, *20*, *25*). The anomaly indicated with the rectangle in Fig. 2A is caused by highland materials within Goldschmidt crater (*4*).

The interval when the Moon is in the Earth's magnetosphere, determined by the Kaguya plasma and magnetic field instruments (*43-45*) data on February 2009, is from 2009-02-07/10:00 to 2009-02-11/05:00 (3.8 days). It is found that the lunar OH/$H_2O$ abundance when inside the Earth's magnetosphere (gray shading in Fig. 2 A-B) remains nearly the same level as those when the Moon is exposed to solar wind.

It is well known that when the Moon passes through the Earth's magnetosphere, solar wind incident on the Moon vanishes. To investigate the correlation between lunar surface hydration and the incident ion energy flux, we computed the average differential ion energy spectrum and ~ 1 keV (typical hydrogen ion energy in the solar wind) median/inter-quartile values of ion energy flux incident on the Moon as a function of lunar phase (0 defines "full moon" and -14/14 indicates "new moon"). This is done using the data available from the Acceleration, Reconnection, Turbulence, and Electrodynamics of the Moon's Interaction with the Sun (ARTEMIS) (*46*, *47*) Electrostatic Analyzer (ESA) instruments (1-25000 eV) (*48*) from September, 2011 to October, 2016 (As shown in Fig.2 C-D). In the computation, we excluded the periods when the probe traversed the lunar nightside in order to avoid the influence of the lunar wake in which the solar wind is shielded. We find that the incoming ~1 keV ion flux when within the Earth's magnetosphere is about two orders of magnitude lower than that from the solar wind Fig. 2 (C-D).

**Discussion:** As we mentioned in the previous sections, the solar wind is thought to be a primary source for lunar global surficial OH/$H_2O$ (*3*, *4*, *12*). The above results have shown that the hydrogen ion energy flux at ~1 keV in the Earth's magnetosphere is two orders lower than that in the solar wind. Therefore, it appears contradictory that the OH/$H_2O$ abundance at lunar polar regions remains at the same level in the two circumstances. To resolve this issue, it is worthwhile to discuss OH/$H_2O$ formation and loss processes in lunar surface minerals.

The bombardment of soil grains on the lunar surface by solar wind protons can produce vacancies at crystal lattice sites, leading to the formation of amorphous rims (*49*). After this sputtering process, the implanted solar wind can be temporarily trapped by vacancies at broken chemical bonds of oxygen, forming individual OH/$H_2O$ (*20*). However, the formed OH/$H_2O$ is unstable; if the strength of the interatomic attractive potentials between the solar wind hydrogen and the regolith oxides (activation energy) is not large enough to cause trapping, the hydrogen atoms will loiter within minerals until diffusing back out into space. The solar wind hydrogen retention time depends on the activation energy and lunar surface temperature; the implanted hydrogen would effectively dwell within lunar minerals in cold regions with larger activation energy (*24*, *50*, *51*).

When the Moon passes through the Earth's magnetosphere, the bulk of the solar wind is shielded. If there were no other sources providing additional OH/$H_2O$ at the same time, its abundance should decrease with time due to the diffusive losses caused by thermal motion. The retained hydrogen in lunar minerals can be calculated using the analytical approach given by (*52*), in which the surface temperature at polar regions is 280 K (measured by the $M^3$ data) and the length of time over which the surface is warmed in the calculation is 3.8 days (determined by the Kaguya MAP-PACE and MAP-LMAG measurements). As shown in Fig.3A, the implanted H with activation energy above ~0.7 eV is retained, while those with activation energy $U < 0.65$ eV tend to escape, resulting in an overall reduction in retained H to 46% of the initially implanted level (Fig.3B). Therefore, the observed reduction in solar wind hydrogen flux while the Moon lies within the Earth's magnetosphere should produce a significant reduction in surface water—if the loss rate discussed above is reliable, and there are no other hydrogen sources available. The fact that a reduction in surface OH/$H_2O$ is not observed implies that possible sources other than solar wind might exist, e.g., large water reservoirs at the lunar polar regions and/or the Earth wind.

First, if there are very large water reservoirs at the polar regions, the surface OH/$H_2O$ would hardly be affected by the lack of solar wind over a few days, and then, any small changes would be attributable to migration/diffusion of the hydrated molecules. For instance, a high abundance of water was found within the plume impact on Cabeus crater by the Lunar Crater Observation and Sensing Satellite (LCROSS) mission (*5*), and orbital spectral albedo observations suggest that the water ice layers in these craters are due to the extreme cold temperature (*7*, *53*). However, the only significant reservoirs known are highly localized and sparsely scattered, occurring within permanently shadowed regions inside craters. Polar water could transport from reservoirs over a wider region and lower latitudes via micrometeoroid impact vaporization and solar wind sputtering. However, the sparsity of these sources is such that the source intensity from this process is too low to fully account for the infrared (IR) observations (*54*).

Alternatively, particles from the Earth's magnetosphere could be a primary source in the absence of solar wind. *Starukhina et al.* (*13*) proposed that the Earth's magnetosphere can provide hydrogen atoms for the lunar regolith in permanently shadowed regions because of the thermal distribution of magnetosphere plasmas; *Lucey et al.* (*55*) also suggested that ions from the Earth's magnetosphere can be a source of lunar volatiles. *Harada et al.* (*56*) found that backscattered hydrogen energetic neutral atom (ENA) flux from the Moon in the Earth's magnetosphere plasma sheet is roughly on the same order of magnitude as that in the solar wind, which indicates a significant amount of Earth wind can reach the Moon.

As shown in the above results, Earth wind proton energy flux at ~1 keV is about two orders of magnitude lower than that from the solar wind (Fig.2C-D), but the OH/$H_2O$ abundance at the lunar polar regions remains the same level in the two circumstances (Fig.2A-B). This may indicate that protons at other energies may provide a significant contribution to OH/$H_2O$ formation. Protons at higher energies can produce more vacancies, e.g. 30 keV proton generates about eight vacancies per incident ion (*57*), which can trap more hydrogen to form OH/$H_2O$, while the 1 keV protons dominant in the solar wind can only create about two vacancies per ion (*52*). Meanwhile, high energy protons can produce deeper vacancy sites (*28*), which can hinder their diffusive loss. On the other hand, protons at energies lower than 1 keV may be trapped more easily by vacancies. This is because the ion-solid collision cross-section (which describes the probability of transferring energy from ions to a solid), is inversely proportional to the square of ion energy (*58*).

To further clarify the different contributions of protons at different energies to the observed surficial OH/$H_2O$ distribution, it is necessary to discuss the incident proton fluxes in the solar wind and the Earth wind in detail. Therefore, in Fig. 4, we plot differential proton energy fluxes in the solar wind and in the Earth wind using the averaged 5-year ARTEMIS ESA data. Two groups of intersection points are obtained: the lower energy intersection points around 325-430 eV, and the higher intersection around 2.5-4 keV. The average differential proton energy fluxes with energy values less than 325 eV or more than 4 keV is larger in the Earth wind than that in the solar wind; and those with energy value between 430 eV-2.5 keV is larger in the solar wind. Compared to the solar wind, the Earth wind protons with energies above 4 keV can therefore produce many more vacancies, and protons with energies below 325 eV can be more easily trapped by these vacancies and generate more OH/$H_2O$. In other words, even though Earth wind proton energy flux at ~1 keV is about two orders of magnitude lower than solar wind, the protons at these two energy ranges may provide a more efficient contribution to the formation of OH/$H_2O$ inside the Earth's magnetosphere, which may result in the same order of lunar OH/$H_2O$ abundance inside/outside the Earth's magnetosphere. On the other hand, if lunar hydration were caused primarily by protons in some specific "optimal" energy ranges, Fig. 4 may give some clues to the constancy of OH/$H_2O$ abundance at the polar regions over one lunation. The existence of these optimal interaction energies may be in the two groups of intersection points, which however needs to be studied by further theoretical simulations and laboratory experiments.

In addition, heavy ions such as oxygen and nitrogen ions in the Earth wind might also contribute to the OH/$H_2O$ production process. Terrestrial oxygen ions could flow out primarily from the polar ionosphere, and escape to the magnetosphere (*34, 35*). During some special periods, e.g., in geomagnetic active times, the oxygen ion flux can be enhanced and even sometimes comparable to the proton flux (*36*). A recent study by the Kaguya spacecraft shows that biogenic terrestrial high energy oxygen ions (1-10 keV) carried by the Earth wind could transport to the Moon and be implanted into the surface of the lunar regolith (*41*). Then oxygen ions in the Earth wind might provide oxygen species to the OH/$H_2O$. What's more, irradiation by heavier ions with greater incident energies would be more efficient to the sputtering process that may add more vacancies (*59*). Thus, the high energy heavy ions in the Earth wind would produce more vacancies in lunar minerals by sputtering processes, which can hold more implanted protons to form OH/$H_2O$.

In a case study, Hendrix et al. (*60*) found that the OH/$H_2O$ abundance is nearly the same in two different locations inside/outside Earth's magnetosphere at a mid-latitude region using the Lunar Reconnaissance Orbiter (LRO) Lyman-Alpha Mapping Project (LAMP) data. On the contrary, using the $M^3$ OP2C data from a statistical study, Li et al. (*61*) suggested that the shielding effect of Earth's magnetosphere on the formation of the lunar surficial water is pronounced at latitudes 60°- 75° S, although this effect is obscured at lower latitudes due to compositional variations of the regolith. From their figures, it appears that the shielding effect of the Earth's magnetosphere tends to be weaker above 65°S, therefore this does not conflict with our observations at the polar regions.

The different variations of lunar OH/$H_2O$ abundance inside/outside of the Earth's magnetosphere at different latitudes may be related to both loss and formation processes. As mentioned above, the thermal diffusion loss rate is higher at lower latitudes because of its higher temperature (*52*). However, our calculation above shows that the loss rate at high latitudes is significant and therefore some formation process must be involved to maintain the same level of water abundance. In essence, solar wind protons have a beam-like velocity distribution with a narrow thermal spread centered at the solar wind bulk velocity. However, the protons in the Earth's magnetosphere, especially in the plasma sheet, have a broad velocity distribution with a wide thermal spread (*56*, *62*) and a larger portion can reach the lunar surface (*63*), which may result in hydrogen delivery to high-latitude regions more effectively. At low latitudes the diffusive loss rate is higher and the incident Earth wind supply may not be sufficient, which may result in the observed magnetotail shielding effect (*64*). This should be investigated in future work.

Besides the Chandrayaan-1 $M^3$, the LAMP onboard the LRO has obtained a large amount of UV spectral data covering the entire lunar surface from 2009 to the present. This provides an opportunity for comprehensive studies on the spatial distribution and temporal variations of lunar water (*22*), and can give more evidence on its formation mechanisms. In addition, the Chinese Chang'E-5 lunar sample-return mission is planned to be launched in 2019. It has an infrared spectrometer with spectral range up to 3 μm, which can also help us to study lunar surface water (OH/$H_2O$). ARTEMIS will concurrently provide accurate and high-resolution plasma data around the Moon in the solar wind and in the magnetosphere.

**References and Notes:**

1. A. E. Saal *et al.*, *Nature* **454**, 192 (2008).
2. R. N. Clark, *Science* **326**, 562 (2009).
3. J. M. Sunshine *et al.*, *Science* **326**, 565 (2009).

4. C. M. Pieters *et al.*, *Science* **326**, 568 (2009).
5. A. Colaprete *et al.*, *Science* **330**, 463 (2010).
6. J. W. Boyce *et al.*, *Nature* **466**, 466 (2010).
7. G. R. Gladstone *et al.*, *Science* **330**, 472 (2010).
8. J. P. Greenwood *et al.*, *Nat. Geosci.* **4**, 79 (2011).
9. E. H. Hauri, T. Weinreich, A. E. Saal, M. C. Rutherford, J. A. Van Orman, *Science* **333**, 213 (2011).
10. H. Hui, A. H. Peslier, Y. Zhang, C. R. Neal, *Nat. Geosci.* **6**, 177 (2013).
11. K. Watson, B. C. Murray, H. Brown, *J. Geophys. Res.* **66**, 3033 (1961).
12. J. R. Arnold, *J. Geophys. Res.* **84**, 5659 (1979).
13. L. Starukhina, Y. Shkuratov, *Icarus* **147**, 585 (2000).
14. F. M. McCubbin *et al.*, *Proc. Natl. Acad. Sci. U.S.A.* **107**, 11223 (2010).
15. A. E. Saal, E. H. Hauri, J. A. Van Orman, M. J. Rutherford, *Science* **340**, 1317 (2013).
16. R. E. Milliken, S. Li, *Nat. Geosci.* **10**, 561 (2017).
17. R. R. Keays *et al.*, *Science* **167**, 490 (1970).
18. J. J. Barnes *et al.*, *Nat. Commun.* **7**, 11684 (2016).
19. E. J. Zeller, L. B. Ronca, P. W. Levy, *J. Geophys. Res.* **71**, 4855 (1966).
20. T. B. McCord *et al.*, *J. Geophys. Res.* **116**, E00G05 (2011).
21. G. Y. Kramer *et al.*, *J. Geophys. Res.* **116**, E00G18 (2011).
22. A. R. Hendrix *et al.*, *J. Geophys. Res.* **117**, E12001 (2012).
23. A. R. Poppe, J. S. Halekas, G. T. Delory, W. M. Farrell, *J. Geophys. Res.* **117**, A09105 (2012).
24. W. M. Farrell, D. M. Hurley, M. I. Zimmerman, *Icarus* **255**, 116 (2015).
25. S. Li, R. E. Milliken, *Sci. Adv.* **3**, e1701471 (2017).
26. C. Wöhler, A. Grumpe, A. A. Berezhnoy, V. V. Shevchenko, *Sci. Adv.* **3**, e1701286 (2017).
27. J. L. Bandfield, M. J. Poston, R. L. Klima, C. S. Edwards, *Nat. Geosci.* **11**, 173 (2018).
28. Z. Djouadi *et al.*, *Astr. Astrophys.* **531**, A96 (2011).
29. A. S. Ichimura, A. P. Zent, R. C. Quinn, M. R. Sanchez, L. A. Taylor, *Earth Planet. Sci. Lett.* **345-348**, 90 (2012).
30. Y. Liu *et al.*, *Nat. Geosci.* **5**, 779 (2012).
31. M. J. Schaible, R. A. Baragiola, *J. Geophys. Res.* **119**, 2017 (2014).

32. J. A. Hundhausen, *Rev. Geophys.* **8**, 729 (1970).
33. S. P. Christon *et al.*, *Geophys. Res. Lett.* **21**, 3023 (1994).
34. K. Seki, R. C. Elphic, M. Hirahara, T. Terasawa, T. Mukai, *Science* **291**, 1939 (2001).
35. Q. G. Zong *et al.*, *J. Geophys. Res.* **106**, 25541 (2001).
36. S. Y. Fu, Q. G. Zong, B. Wilken, Z. Y. Pu, *Space Sci Rev* **95**, 539 (2001).
37. H. Hasegawa *et al.*, *Nature* **430**, 755 (2004).
38. Q. Q. Shi *et al.*, *Nat. Commun.* **4**, 1466 (2013).
39. A. R. Poppe, M. O. Fillingim, J. S. Halekas, J. Raeder, V. Angelopoulos, *Geophys. Res. Lett.* **43**, 6749 (2016).
40. R. P. Lin *et al.*, *Space Sci Rev* **71**, 125 (1995).
41. K. Terada *et al.*, *Nat. Astron.* **1**, 26 (2017).
42. S. Li, R. E. Milliken, *J. Geophys. Res.* **121**, 2081 (2016).
43. Y. Saito *et al.*, *Earth Planets Space* **60**, 375 (2008).
44. Y. Saito *et al.*, *Space Sci Rev* **154**, 265 (2010).
45. H. Tsunakawa *et al.*, *Space Sci Rev* **154**, 219 (2010).
46. V. Angelopoulos, *Space Sci Rev* **165**, 3 (2011).
47. D. G. Sibeck *et al.*, *Space Sci Rev* **165**, 59 (2011).
48. J. P. McFadden *et al.*, *Space Sci Rev* **141**, 277 (2008).
49. L. P. Keller, D. S. McKay, *Geochimica et Cosmochimica Acta* **61**, 2331 (1997).
50. L. Starukhina, *J. Geophys. Res.* **106**, 14701 (2001).
51. L. Starukhina, *Water on the Moon: What Is Derived from the Observations?* (Springer Berlin Heidelberg, 2012), pp. 57-85.
52. W. M. Farrell, D. M. Hurley, V. J. Esposito, J. L. McLain, M. I. Zimmerman, *J. Geophys. Res.* **122**, 269 (2017).
53. S. Li *et al.*, *Proc. Natl. Acad. Sci. U.S.A.* **115**, 8907 (2018).
54. W. M. Farrell *et al.*, *Planet. Space Sci.* **89**, 15 (2013).
55. P. G. Lucey, *Elements* **5**, 41 (2009).
56. Y. Harada *et al.*, *J. Geophys. Res.* **119**, 3573 (2014).
57. D. Fink *et al.*, *Applied Physics A* **61**, 381 (1995).
58. M. Nastasi, J. W. Mayer, J. K. Hirvonen, *Ion-Solid Interactions: Fundamentals and Applications.*, (Cambridge Univ. Press, Cambridge, U. K., 1996).
59. A. F. Barghouty, F. W. Meyer, P. R. Harris, J. H. A. Jr, *Nuclear Inst & Methods in Physics Research B* **269**, 1310 (2011).

60. A. R. Hendrix *et al*., 48th Lunar and Planetary Science Conference, abstract 2149, The Woodlands, Texas, 20 to 24 March 2017.

61. S. Li, P. G. Lucey, T. M. Orlando, 49th Lunar and Planetary Science Conference, abstract 1575, The Woodlands, Texas, 19 to 23 March 2017.

62. W. C. Feldman *et al*., *J. Geophys. Res.* **106**, 23231 (2001).

63. D. A. Hardy, H. K. Hills, J. W. Freeman, *Geophys. Res. Lett.* **2**, 169 (1975).

64. E. Cho, Y. Yi, J. Yu, I. Hong, Y. Choi, *J. Geophys. Res.* **123**, 2110 (2018).

**Acknowledgments:** The authors express special thanks to W. M. Farrell and O. J. Tucker for providing detailed calculation of evolution of distribution of activation energy. We also gratefully thank constructive discussions with K. K. Khurana, S. Li and M. G. Kivelson. We thank all the members of the Chandrayaan-1 $M^3$, Kaguya MAP-PACE and MAP-LMAG, and ARTEMIS ESA instrument teams. **Funding:** This research was supported by the Key Research Program of the Chinese Academy of Sciences (Grant NO. XDPB11), the National Natural Science Foundation of China (Grants 41574157, 41774153, 41773065, 41490634, and 41628402), the Specialized Research Fund for State Key Laboratories; **Competing interests:** Authors declare no competing interests; and **Data and materials availability:** The $M^3$ and Kaguya data were downloaded from the Planetary Data System and the SELENE Data Archive. The ARTEMIS data are available at http://artemis.ssl.berkeley.edu.

**Supplementary Materials:**

Materials and Methods

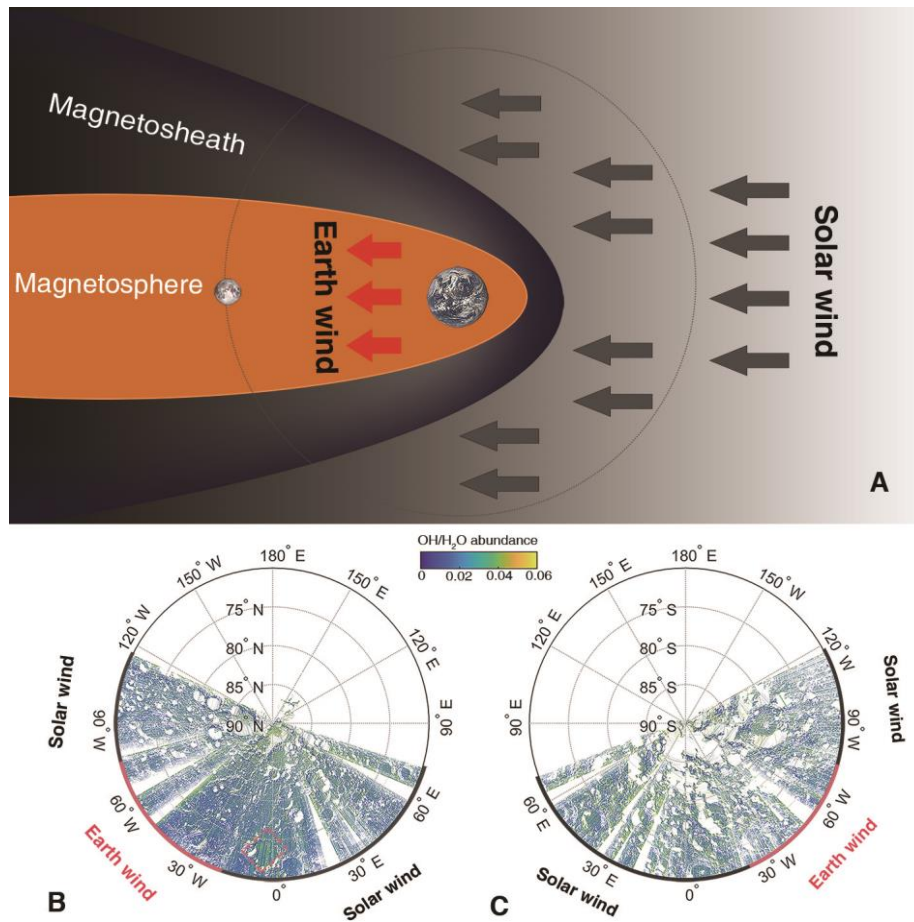

**Fig. 1. (A) Lunar orbit in a lunation**, indicating that during 3-5 days of every lunation, the Moon lies within the Earth's magnetosphere, in which the solar wind is shielded. **(B-C) Overview of lunar OH/$H_2O$ abundance (represented here by strength of 2.8 μm absorption feature) in the polar regions derived from the $M^3$ observations from 2009-Jan-26 to 2009-Feb-23.** Red and black bars mark the duration of the Moon in the Earth wind and solar wind, respectively. In general, we can find that there were no significant differences in OH/$H_2O$ abundance in the Earth wind and solar wind. The red rectangle indicates the Goldschmidt crater, in which lunar OH//$H_2O$ abundance is anomalously high due to its material composition.

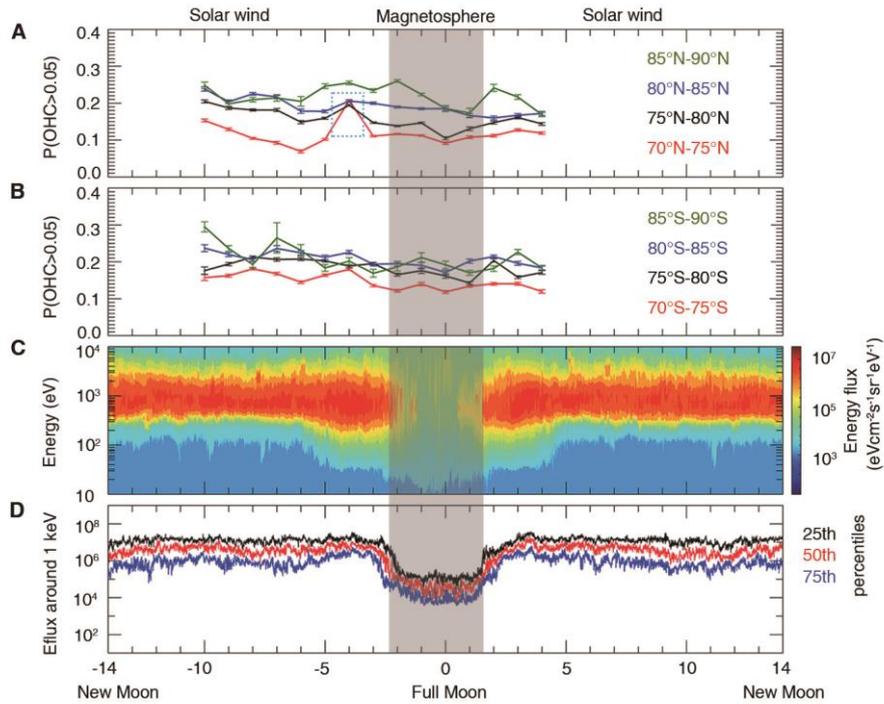

**Fig. 2. OH/H$_2$O abundance at lunar polar regions during a lunation from 26 Jan 2009 to 23 Feb 2009 and ion energy flux distribution averaged over five-year observations. (A-B)** The probability of 2.8 μm absorption depth between 0.05 and 0.2 represents the OH/H$_2$O abundance level for each 5° latitude and 24 Earth hour bin. The zero epoch in top two panels indicates full-moon time during the M$^3$ observation. The error is calculated by $\varepsilon = \frac{n}{N} * (\frac{\sqrt{n}}{n} + \frac{\sqrt{N}}{N})$, where *n* is the number of pixels with an absorption depth greater than 0.05, and *N* is the number of pixels with an absorption depth less than 0.2 in each bin. The error bar shown in the figure is scaled by a factor of 10. We can find that the lunar surface OH/H$_2$O abundance increases with latitude toward the polar regions, which is consistent with previous studies (*4*, *20*, *25*), and the anomaly indicated with blue dashed line rectangle is caused by highland materials within Goldschmidt crater. The gray-shaded area marks the duration of the Moon in the Earth's magnetosphere determined by Kaguya observations. From these two panels it is clearly seen that the OH/H$_2$O abundance inside the Earth's magnetosphere remains nearly the same level as those when the Moon is exposed to solar wind. **(C-D)** A superposed epoch analysis of (C) ion spectrogram and (D) ion energy flux around 1 keV as a function of lunar phase using the five-year ARTEMIS data from Sep 2011 to October 2016, where the zero epoch in the two bottom panels indicates the full-moon times, for five years of ARTEMIS observations. The red solid line represents the median value of energy flux, and the black and blue solid lines represent the upper/lower quartiles. The ion energy distribution in the magnetosphere becomes broader and the ion energy flux around 1 keV is about two orders of magnitude lower than that in the solar wind. This figure suggests that in the Earth wind the protons at other energies and heavy ions may provide significant contribution to the OH/$H_2O$ formation, see details in the text.

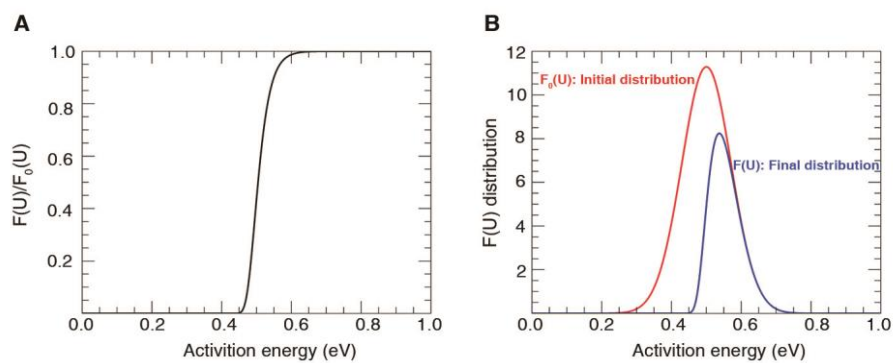

**Fig 3. Temporal evolution of implanted hydrogen distribution with activation energy when the sources are turned off. The results are calculated using the model described in (*52*), with T = 280 K and duration Δt = 3.8 days. (A) The ratio between the final and initial H distribution.** The implanted H with activation energy above ~0.7 eV is retained, while those with U < 0.65 eV tend to escape. **(B) Initial (red) and final H (blue) distributions**. The overall retained H for all activation energies is down to 46% of those initially implanted. Therefore, possible sources other than solar wind should exist in the Earth's magnetosphere.

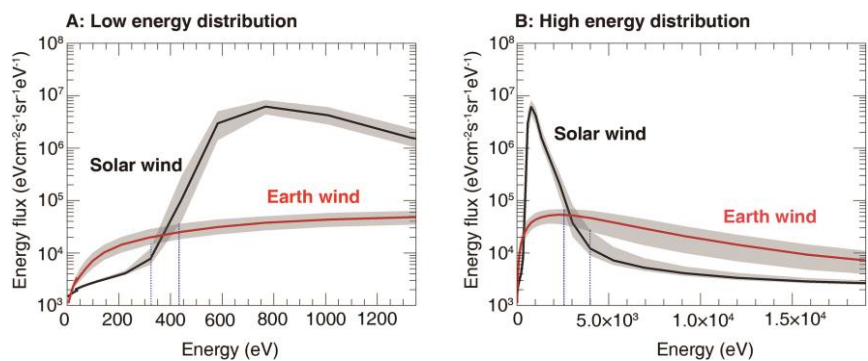

**Fig 4. Differential ion energy fluxes for (A) low and (B) high energy ranges in the solar wind and in the magnetosphere from averaged 5-year ARTEMIS ESA data.** The two groups of intersection points are around 325-430 eV and 2.5-4 keV. The average differential proton energy fluxes with energy value less than 325 eV or more than 4 keV are larger in the Earth wind than those in the solar wind. We can therefore expect that the Earth wind protons at energies above 4 keV produce more vacancies, while protons below 325 eV can be easily trapped (because of their lower energy) by these more abundant vacancies compared to solar wind. Therefore, even though Earth wind proton energy flux at ~1 keV is about two orders of magnitude lower than solar

wind, the protons at these two energy ranges could be a significant contribution to the formation of OH/$H_2O$ inside the Earth's magnetosphere. On the other hand, if lunar hydration were caused primarily by protons in some specific "optimal" energy bands, the existence of these optimal interaction energies may be in the two groups of intersection points.

**Materials and Methods**

The $M^3$ imaging spectrometer onboard the Chandrayaan-1 mission measures reflected solar light by the lunar surface, covering visible and near-infrared wavelengths between ~460 nm and 3000 nm. This spectral range contains absorption features near 2.8 to 3 μm caused by OH/$H_2O$ (*4*). We selected the $M^3$ data within a lunation from January 26, 2009 to February 23, 2009, while the optical period is OP1B in cold thermal operating conditions. During this time interval the Moon stayed in the Earth's magnetosphere for several days around February 9, 2009 (the full moon time). Lunar near-infrared (NIR) reflectance spectra is affected by the thermal emission radiation at wavelengths beyond 2 μm (*42*). In order to suppress thermal residuals in the first-order thermal corrections of the $M^3$ data, only the polar regions with latitude ab2ove 70° are selected due to their lower temperatures (~ 280 K). The "bad" pixels with lower and flat reflectance spectra are removed based on their wavelength-reflectance 2D histogram. Then, we calculated the absorption depth around 2.8 μm as an indicator of lunar surface OH/$H_2O$ abundance, as described in (*4*).